\documentclass[runningheads]{llncs}
\usepackage[T1]{fontenc}
\usepackage{graphicx}
\usepackage{booktabs}
\usepackage{epigraph}
\usepackage{subcaption}
\usepackage{xcolor}
\usepackage{hyperref}
\usepackage{xspace}
\usepackage{amssymb}
\usepackage[most]{tcolorbox}
\usepackage[normalem]{ulem}  

\usepackage{hwemoji}

\newcommand{\casestudy}{🔍\xspace}
\newcommand{\bluesky}{🌱\xspace}

\newcommand{\secret}{\ensuremath{s}\xspace}
\newcommand{\secretspace}{\ensuremath{\mathbb{S}}\xspace}
\newcommand{\obs}{\ensuremath{o}\xspace}
\newcommand{\obsspace}{\ensuremath{\mathbb{O}}\xspace}
\newcommand{\train}{\ensuremath{\mathcal{T}}\xspace}

\newcommand{\predictor}{\ensuremath{f}\xspace}
\newcommand{\cp}{\ensuremath{C}}

\begin{document}
\title{Conformal Prediction for Offensive Security}
\author{Giovanni Cherubin} %
\authorrunning{G. Cherubin}
\institute{Microsoft, Cambridge, UK\\
\email{gcherubin@microsoft.com}}
\maketitle              %
\begin{abstract}
Despite its introduction more than a quarter century ago, Conformal Prediction (CP) has seen surprisingly
few applications to the cyber security world thus far.
In particular, we observe that, while CP has been employed as a defensive measure in many recent works,
its use for carrying out attacks (i.e., for \textit{offensive security}) is hard to trace in the literature.
We explore this gap, by presenting initial findings in two key areas
of offensive security: Privacy-Preserving Machine Learning, and
network traffic analysis.

\end{abstract}

\epigraph{To Alex: PhD supervisor, mentor, and friend, on the occasion of his 80th birthday, in recognition of his guiding influence and pioneering work in Machine Learning and Conformal Prediction.}

\section{Introduction}

Cyber Security offers multiple and varied opportunities for problem solving and scientific discovery.
Importantly, security has a direct impact on society; this can lead to success stories
(e.g., it is nowadays possible for two people to communicate in full privacy over the Internet
thanks to end-to-end encryption protocols~\cite{theNewTextSecure}),
as well as unsuccessful ones (e.g., the use of poor anonymization techniques can lead to
user identification~\cite{narayanan2006break}).
Differently from many fields in Computer Science, security deals with adversarial parties;
this forces researchers to continuously
look at the security of a system from two sides, the defender's and the attacker's,
and it encourages a continuous improvement from both.

Offensive research, which aims to improve attacks against computer systems,
is an essential part of cyber security.
First, it helps grounding work on defenses, because of an asymmetry between attacks and defenses:
a defense needs to protect against most (or all) attacks in a threat scenario,
whereas an attack just needs to bypass one defense to be considered successful.
Secondly, offensive research informs the public and product vendors on what attack techniques can be used:
it would be a fallacy to think that reducing research on attacks
is a good way of preventing cyber crime.

Machine Learning (ML), and more recently Conformal Prediction (CP), have demonstrated
a plethora of successful applications to \textit{defensive} security;
here, ``defensive'' refers to using ML or CP for attack prevention or detection.
In particular, numerous defenses have been developed in recent years where CP is the
main basis for protecting a system.
They range from network intrusion~\cite{wechsler2015cyberspace,cherubin2015conformal,dang2023kernel},
and malware~\cite{jordaney2017transcend} detection, to a large body of works evaluating CP's
ability to prevent ML evasion attacks (so-called ``adversarial examples'') and poisoning attacks~\cite{ennadir2023conformalized,messoudi2020deep,gendler2021adversarially,kang2024colep,luo2024game,bao2025enhancing,cherubin2018exchangeability};
more recently, CP has also been investigated as a defense mechanism for privacy-preserving ML~\cite{angelopoulos2022private,balinsky2025conformal}.

On \textit{offensive} security, whilst traditional ML has seen a similarly successful adoption~\cite{cai2012touching,johnson2013users,panchenko2016website,hayes2016k,juarez2014critical,cherubin2017bayes,sirinam2018deep,shokri2017membership,fredrikson2015model,carlini2021extracting,song2001timing,hettwer2020applications,cherubin2019f},
CP has been largely unexploited; in fact,
to the best of our knowledge, no attack to date employs CP in an offensive manner.
This is made even more surprising by the fact that security researchers have historically been
early adopters of new ML techniques for improving attacks;
this was quite evident, for example, in the Website Fingerprinting community, which rapidly adopted
increasingly advanced neural network techniques for improving their attacks~\cite{panchenko2016website,sirinam2018deep,sirinam2019triplet}.

We attribute the lack of CP applications to offensive security research to a simple fact:
CP is not a plug\&play alternative to traditional ML.
Simply, and this is part of its beauty, conformal inference requires a different way of framing the problem.
Far from being an obstacle, we argue that this switch of perspective
opens up many new possibilities for offensive security research.
In this work, we support this claim by introducing new CP-based attacks in two areas of cybersecurity.
Along the way, we identify features that make these attacks particularly well suited for 
CP, such as the ability to generate arbitrarily large datasets and the value of producing
prediction sets rather than point predictions.
These aspects highlight how CP provides unique advantages over conventional ML approaches in offensive security settings.

\section{Preliminaries}
\label{sec:preliminaries}

We describe how traditional ML models (classifiers or regressors) are typically
used for carrying out attacks, and then introduce how CP can be adapted for the same
purpose.

\subsection*{Traditional ML applications to offensive security}
Many security attacks can be modeled as instances of a statistical inference game~\cite{cherubin2019f}:
an adversary tries to predict some secret information $\secret \in \secretspace$
based on side information (or ``observation'') $\obs \in \obsspace$ they have available;
in attacks of this form, we model the relation between $\obs$ and $\secret$ via their joint
distribution $P(\secret, \obs) = P(\secret)\,P(\obs\mid \secret)$.
For example, consider a password guessing attack: the secret \secret is the user’s password,
and the observation $\obs$ could be partial knowledge of the password, system responses to attempted logins (e.g., timing hints from login failures),
or any other signals that help narrow down the correct guess.
Typically, the attacker can further observe (and learn from) examples of the
form $(\obs_i, \secret_i) \in \obsspace \times \secretspace$, which are sampled
from the joint $P(\secret, \obs)$.
The goal of the attacker is to guess the password $\secret$ based on this
knowledge $\obs$.

In this formulation, it is perhaps not surprising that traditional ML has been an extremely successful tool in
carrying out attacks: this setup has the same form of classical statistical learning problems,
(classification or regression), where a learner aims to train a predictor
$\predictor: \obsspace \mapsto \secretspace$ that makes guesses for secret information
$\secret$ based on the side information $\obs$.

More concretely, traditional ML models can be applied to security attacks of this form as follows.
An attacker obtains a training set $\train \in (\obsspace \times \secretspace)^n$ by sampling
from $P(\secret, \obs)$.\footnote{In practice, this sampling usually happens in two steps, by first sampling the prior $P(s)$ and then
the conditional $P(\obs\mid \secret)$.
In general, we assume the attacker can sample from this distribution (oracle access),
but they may not know the distribution itself. 
}
For example, in the password guessing scenario, the attacker could collect a variety of
passwords $s_i$ alongside timing responses or other feedback signals from login attempts ($o_i$).
On this basis, the attacker trains an ML predictor (classifier or regressor) by minimizing some
loss;
e.g., for classification problems (i.e., the secret information $s$ is a categorical value),
the attacker trains:
$$\predictor = \arg\min_\predictor \sum_{\secret, \obs \in \train} I(\predictor(\obs) \neq \secret)\,$$
where $I$ is the indicator function.
The attacker can then use $\predictor$ for carrying out the attack, to make a prediction
$\secret' = \predictor(o)$ for the true secret $\secret$.

\subsection*{The Conformal Prediction way to offensive security}

Conformal Prediction (CP) can be readily applied to security attacks of the form
described above, although it often requires reframing the inference problem.
As we argue in this chapter, this adjustment is not a drawback but rather an opportunity
to devise novel and often more informative attacks.

In what follows, we adopt a simplified notation for CP and refer to the literature for a more formal introduction~\cite{vovk2005algorithmic}.
We denote by
$\cp^\alpha_{\train}: \obsspace \mapsto \mathcal{P}(\secretspace)$
a conformal predictor, where $\train \in (\obsspace \times \secretspace)^n$ is a
training dataset and $\mathcal{P}(\cdot)$ is the power set.
Informally, $\cp^\alpha_{\train}(\obs)$ produces a subset of $\secretspace$ that
contains all secrets deemed plausible for the new observation $\obs$;
for brevity, we omit the underlying nonconformity function, which is typically
based on a traditional ML model (e.g., a regressor or probabilistic classifier).
A conformal predictor is defined for a significance level $\alpha \in [0,1]$
and it assumes that $\train$ and each new example $(\obs, \secret)$ come from an
exchangeable distribution. Under these conditions,
CP provides the following \emph{validity} guarantee:\footnote{Here, the probability space is the realization of the exchangeable distribution, and
the guarantee is marginal: it holds true on average for new samples
$\train \cup \{(\obs, \secret)\}$, but not conditionally on each $(\obs, \secret)$.}

$$P(\secret \notin \cp^\alpha_{\train}(\obs)) \;\leq\; \alpha.$$

Integrating conformal prediction (CP) into an attack that relies on a traditional ML model $\predictor$
is often straightforward: one can simply “wrap” CP around
$\predictor$~\cite{vovk2005algorithmic,papadopoulos2008inductive,angelopoulos2021gentle}.
However, there are scenarios where such a direct approach brings limited benefits,
while CP’s emphasis on rethinking the problem can lead to new interesting and impactful attacks (\autoref{sec:mia}).

For an attacker using CP, the most immediate consequence is that, rather than a point prediction,
CP returns a set of labels (the prediction region).
This, as shown in various examples throughout this chapter (\autoref{sec:ppml}-\ref{sec:wf}),
can be very informative to an attacker:

\begin{tcolorbox}[  
  title=Remark,  
  colframe=teal!60!white,  
  colback=teal!10,  
  coltitle=black,  
  fonttitle=\bfseries  
]  
When used for offensive security, CP changes the attacker’s objective
from making a single best guess to producing a set of plausible secrets,
with a guaranteed error rate ($\alpha$).
This can make attacks more efficient,
as it gives the attacker a principled way to focus
on a smaller
set of candidates.
For instance, in a password-guessing scenario, CP might return a relatively small set of
possible passwords, which an attacker can then systematically test;
by focusing on this smaller subset, the attacker significantly reduces the time and computational cost of a full brute-force search.
\end{tcolorbox}

On this aspect, one may wonder if applying CP to security is equivalent to using multi-label ML
methods (i.e., returning the top-$k$ candidate labels from a traditional ML predictor, or returning
all candidates whose confidence exceeds a certain threshold).
A crucial difference is that CP offers a principled solution to the problem
-- namely it does not require setting arbitrary thresholds on the number of candidates to return,
as in CP the size of the prediction set is determined dynamically depending on the underlying probability distribution.

We also observe that CP's validity guarantee comes handy to an attacker:
from the attacker's perspective, there is a risk in
reducing the set of candidate values
to explore when trying to guess a secret,
since it may lead to missing the correct secret entirely.
CP's validity guarantee provides a formal way
to quantify and control this risk directly via $\alpha$.

In the next sections, we study attacks from two different
fields of Cyber Security and how they benefit from using CP.
We split this discussion between case studies (\casestudy{}),
which contain initial experiments that showcase and study the attacks, and
research directions (\bluesky{}), which set out the basis for ideas to explore.

\section{Privacy-preserving ML}
\label{sec:ppml}

The field of Privacy-preserving ML (PPML) is concerned with the privacy of the data that
is used for training an ML model~\cite{shokri2017membership,fredrikson2015model,carlini2021extracting}.
The main question being asked is as follows. Suppose an ML model was trained on private data,
and its predictions (or the trained parameters of the model itself) were released to the public:
what information could be inferred about the private training data by a member of the public,
such that this information could have not been inferred prior to the model's release?
In this sense, PPML has to do with measuring and exploiting the information leakage of an ML model
about its training data.

Numerous PPML attacks and defenses have been proposed over years.
For a broader look, we refer the reader to Salem et al.~\cite{salem2023sok}, who offers
a comprehensive overview and formalization.
Here, we focus on one: reconstruction attacks.

\subsection{\casestudy{} Case Study: Data Record Reconstruction}
\begin{figure}
  \centering
  \includegraphics[width=\textwidth]{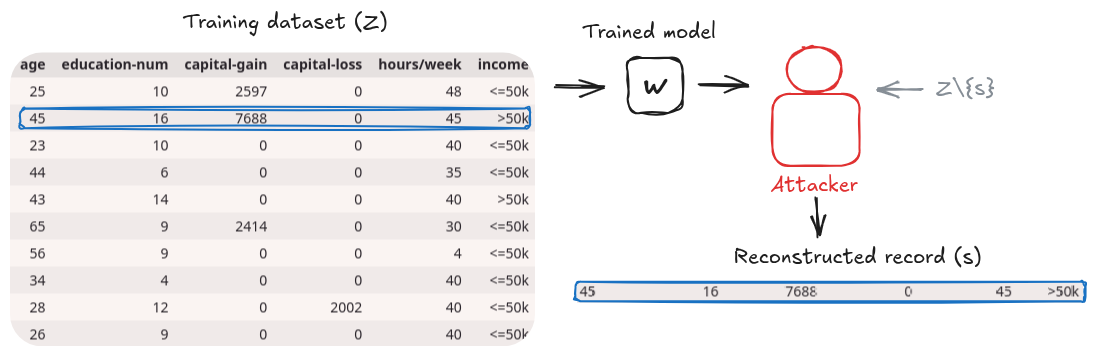}
  \caption{A reconstruction attack.
  The attacker tries to infer the features (and label) of one of the data records that
  were used for training an ML model.
  The attacker has access to the model, and (in this chapter) the entire training dataset minus the data record.
  }
\end{figure}

In reconstruction attacks, an attacker gets access to a trained ML model, the
\textit{target} model,
and they try to
infer (\textit{reconstruct}) the values of one of the training data points that were used
for training the model.
Among PPML attacks, they provide
the highest leakage but can also be among the hardest to carry out~\cite{carlini2021extracting,balle2022reconstructing}.
We explore the potentials of CP when used to carry out a reconstruction attack.

\subsubsection{Problem setup and threat model.}
In reconstruction attacks, an ML model $w$ is trained on a private training dataset
$Z \in (\mathbb{X} \times \mathbb{Y})^N$;
at this stage, it is not important whether the model's task is classification or regression.
It is useful to think of this model in terms of its trained parameters: these are, for example,
the intercept and coefficients if the model is a linear model, or the trained neuron weights in
the case of a neural network.
We shall henceforth interchangeably use $w \in \mathbb{W}$ for referring to the target model and its trained parameters.

An attacker's goal is to infer the value of a data record, $\secret \in Z$, using the following
information:\footnote{Here we use the ``informed attacker'' threat model by Balle et al.~\cite{balle2022reconstructing}.} the model's parameters $w$, and the entire training dataset minus the data record $Z \setminus \{\secret\}$;
we also assume that the attacker knows the hyperparameters of the model, and any randomness source:
that is, they should be able to train a model that is identical to $w$ if they somehow had access to
the full dataset $Z$.
These strong assumptions, while idealised, allow us to explore the theoretical limits
of reconstruction attacks and are consistent with prior work~\cite{balle2022reconstructing}.

According to the notation defined in \autoref{sec:preliminaries}, a reconstruction attacker needs to
infer the value
of a data record (and its label), $\secret \in (\mathbb{X} \times \mathbb{Y}) = \secretspace$, based
on the information they have available: $w \in \mathbb{W} = \obsspace$.\footnote{
  Technically,
  the information available to our attacker is
  $(w, Z \setminus \{\secret\}) \in \obsspace = \mathbb{W} \times (\mathbb{X} \times \mathbb{Y})^{N-1}$,
  but we omit $Z \setminus \{\secret\}$
  for compact notation, and we implicitly assume that the attack is conditioned on a
  specific $Z \setminus \{\secret\}$ and $\secret$ throughout.
}
To carry out the attack,
the attacker trains an ML regressor model $r: \secretspace \mapsto \obsspace$, which enables them
making a guess $s' = r(w, Z \setminus \{\secret\})$ for the true data record $s$;
we measure the goodness of an attack as the distance between $s'$ and $s$ (e.g., $L_2$ distance).
To train the regressor, the attacker generates examples by training ML models
$\{\tilde{w}_i\}_{i=1}^k$ based on datasets $\{Z \setminus \{z\} \cup \{\tilde{s}_i\}\}_{i=1}^k$, where
$\tilde{s}_i \in (\mathbb{X} \times \mathbb{Y})$ are random samples from some
distribution\footnote{In our experiments, we use the empirical distribution
of the set $Z \setminus \{z\}$.}.
We observe that, in principle, this allows generating \textit{arbitrarily many} training examples for
training the attacker's regressor.

To the uninitiated, it may be confusing that there are two ML models at play here:
the target model $w$, which was trained by a data curator, and the attacker's model $r$,
which is used for carrying out the attack.
This fact should not surprise: ML models are very efficient for performing security attacks~\cite{cherubin2019f},
and it is a mere coincidence that the target of the attack in this case involves an ML model.

\subsubsection{CP for reconstruction.}

Reconstruction attacks apply a traditional ML regressor $r$ for inferring a data record $\secret$
from the original training set $Z$ based on the target model's parameters $w$.
Reformulating the attack for an adversary using CP is straightforward:
they can perform CP regression, which outputs a prediction region, rather than a point prediction.
In turn, CP gives a reconstruction attacker prediction bands, which are arguably more informative
than a point prediction when trying to infer a person's private information:
the attacker might use the confidence bands as a signal, to be combined
with information from other sources (e.g., OSINT~\cite{block2024long}), to cause a more impactful attack.

\subsubsection{Experimental setup.}

\paragraph{Target model training.}
In this experiment, the target model is a Logistic Regression classifier trained on the Adult dataset~\cite{adult_2} for the typical classification task: to
predict whether the income of a person in the dataset is above or below 50k (\texttt{income});
the prediction is based on these features: \texttt{age}, \texttt{education-num}, \texttt{capital-gain}, \texttt{capital-loss}, \texttt{hours-per-week}.
We train 5,000 target models; each model is trained on the dataset $Z = Z' \cup \{s_i\}$, where
$Z'$ is the Adult dataset's training set, whereas $s_i$ are points taken from the Adult dataset's test split.
The goal of the reconstruction attack is to reconstruct each $s_i$ based on the respectively trained models.
We also repeat the experiment with Logistic Regression target
models that are trained with Differential Privacy (DP) guarantees~\cite{dwork2006differential},
implemented via \texttt{diffprivlib}~\cite{diffprivlib};
for an adequate DP parameter $\varepsilon$, this defense mechanism ensures that models are robust against reconstruction attacks, thanks to
DP's formal guarantees;
we select $\varepsilon = 5$, which in this case
is a sufficient level of protection to counter reconstruction attacks~\cite{balle2022reconstructing}.

\paragraph{Attacker's model training.}
We employ Conformalized Quantile Regression (CQR) by Romano et al.~\cite{romano2019conformalized}, which
enables tailoring coverage bands to individual data points, and overall improves
CP's performance in regression settings.
We use gradient boosting (LightGBM~\cite{ke2017lightgbm}) to build the underlying nonconformity measure; the implementation for this experiment is based on the MAPIE library~\cite{Cordier_Flexible_and_Systematic_2023}.
For simplicity, we run a separate attack for each feature (and for the label) of the
data record to reconstruct;
the attack's output is obtained by running each of the regressors independently,
and then joining their outputs.
For training the regressors, we generate a dataset of 100k data points; each of them is
a target model $\tilde{w}_i$ trained on $Z' \cup \{\tilde{s}_i\}$, where $Z'$ is the Adult dataset's training
split, and $\tilde{s}_i$ is a random point sampled from the empirical distribution of $Z'$.
We sample $\tilde{s}_i$ from the empirical distribution of $Z'$ to simulate the process of drawing a random
person’s data and then training the corresponding model $\tilde{w}_i$, thus approximating the joint distribution of $(s, w)$.

\paragraph{Performance metrics.}
We evaluate the CP-based attack in terms of its coverage and the size of its prediction regions.
For the point predictor (LightGBM) used to define the nonconformity measure,
we measure the success of the attack by computing the $L_2$ distance between the guessed record $\secret'$ and the true record $\secret$. We also include a naive baseline attack that simply outputs the median value for each feature in $Z \setminus \{s_i\}$, allowing us to see how much an informed attack improves over one that ignores the target model entirely.

\begin{table}
\centering
\caption{
In reconstruction attacks, an entire data record is reconstructed (including its label),
based on having access to an ML model trained on it.
We look at two cases: one where the adversary attacks
a standard Logistic Regression model, without any privacy mechanism in place;
one where a Differential Privacy (DP) mechanism is applied to the Logistic Regression model.
This table shows the results over a test set of 5k examples.
CP's coverage guarantees (for $\alpha=0.1$) are met, and we observe that the prediction sets are
small relatively to the features' range.
}\label{tab:recon}
\vspace{0.5em}
\begin{tabular}{lrrrrrr}
 & Age & Education & Gain & Loss & Hours & Label \\
\toprule
\textbf{No privacy}\\
Coverage & 0.91 & 0.91 & 0.98 & 0.99 & 0.91 & 1.00 \\
Average Prediction Size & 15.32 & 2.67 & 828.86 & 24.23 & 10.91 & 0.12 \\
Prediction $L_2$ norm & 4.69 & 1.18 & 3146.16 & 60.90 & 8.73 & 0.05 \\
Baseline improvement & 0.67 & 0.53 & 0.62 & 0.85 & 0.30 & 0.89 \\
\toprule
\textbf{DP ($\varepsilon=5$)}\\
Coverage & 0.90 & 0.92 & 0.95 & 0.95 & 0.92 & 1.00 \\
Average Prediction Size & 44.58 & 8.81 & 4899.16 & 257.55 & 43.43 & 1.00 \\
Prediction $L_2$ norm & 14.12 & 2.55 & 8364.72 & 411.70 & 12.39 & 0.49 \\
Baseline improvement & 0.00 & 0.00 & 0.00 & 0.00 & 0.00 & 0.00 \\
\toprule
\end{tabular}
\end{table}

\subsubsection{Results.}

\autoref{tab:recon} shows the performance of the CP-based attack;
the reconstruction success metrics are presented individually
for each feature (and for the label) of the data records.
In the no-privacy setting
(i.e., the target model is a Logistic Regression classifier trained without any privacy protection), CP achieves small prediction sets
for each of the features and labels;
furthermore, the point predictions of its underlying classifier
indicate a remarkable advantage over the naive baseline, showing that
the attack is effective.
When a DP privacy-preserving mechanism is applied (with DP $\varepsilon=5$), the attack is unsuccessful
as expected.
In both cases, the CP coverage of $\alpha=0.1$ is attained.

\begin{figure*}
    \centering
    \includegraphics[width=\linewidth]{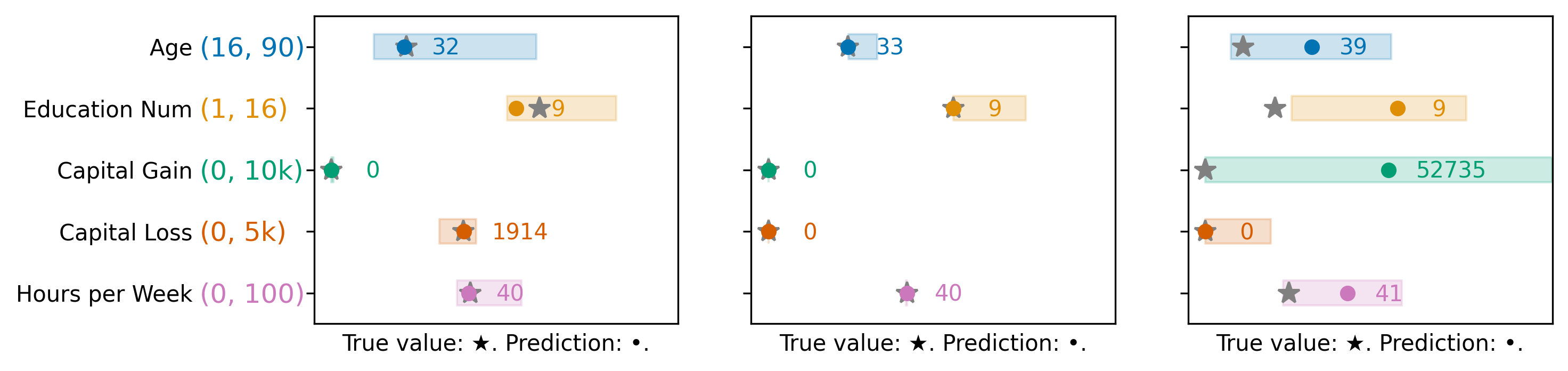}
    \caption{Cherry-picked reconstruction examples: the first one from the left shows perfect validity; in the second one, some of the reconstructed values are on the bands' edges; the third one shows the largest deviation from validity that we observed in our experiments for the feature ``Capital Gain''.}
    \label{fig:reconstructed}
\end{figure*}

\autoref{fig:reconstructed} shows three reconstruction examples, cherry-picked to illustrate
the variety of outcomes we observed.
The first two examples show valid predictions for each feature.
In the third example, the coverage interval
for ``Capital Gain'' falls out of the prediction region, illustrating a case where CP’s
validity fails.
This highlights a disadvantage of CP: because CP's validity guarantee does not hold conditionally
for an example, but on average, there will be individual examples (for $\alpha > 0$) for which
validity is not met.

\subsubsection{Discussion.}
Results confirm that CP is suitable for tackling reconstruction attacks, and it is
a promising avenue for PPML attacks in general.
Its main advantage is clear: an attacker can use CP to get an informative coverage band for their
prediction, which they can combine with further information from other attacks.
This encourages more security research to take into consideration CP
as part of the threat model.
We highlight two characteristics that make reconstruction (as well as several other attacks)
especially suitable for CP:

\begin{tcolorbox}[  
  title=Learnings,  
  colframe=teal!60!white,  
  colback=teal!10,  
  coltitle=black,  
  fonttitle=\bfseries  
]
\noindent
\textbf{Virtually unlimited training examples.}
In some attacks, the attacker can generate as many training examples as they like
For example, in the case of reconstruction attacks, the attacker can generate arbitrarily
many $(\tilde{s}_i, \tilde{w}_i)$ training pairs, by sampling $\tilde{s}_i$ from a known distribution
and fitting the respective model $\tilde{w}_i$.
This makes it possible to use methods such as split CP and CQR, which are typically less data-efficient
but much more computationally efficient than full CP.

\noindent
\textbf{Data is IID.}
In many security attacks, the data available to an attacker for training and in
evaluation can be realistically assumed to come
as independent samples from the same distribution.
In our reconstruction attack, we similarly assume that the target model $w$ is drawn from the same distribution as the training
examples we generate, allowing conformal prediction to retain its validity guarantees.
\end{tcolorbox}

\subsection{\bluesky{} Research direction: Membership Inference attacks}

\label{sec:mia}

The field of PPML has a lot of varied attacks, and we foresee CP being significant for carrying out
many of them.
Some attacks are trivial to adapt to the CP framework; for example, in attribute inference,
where an attacker reconstructs part of a data record, one can readily apply our learnings from the
reconstruction attack described above.
On the other hand, we observe that CP may induce new research directions for other attacks,
such as membership inference.
We propose two new attacks that leverage CP for membership inference.

In traditional membership inference attacks, an attacker gets access to a model trained on private dataset $Z$,
and to a data record $z$, and their goal is to figure out if $z \in Z$ (i.e., if the record
was a \textit{member} of the training data).
For an attacker, this resolves in a binary classification problem, whose goal is to predict the
indicator $I(z \in Z)$.
This makes the direct application of CP classification to this setting straightforward, but
possibly uninsightful.
There are, however, two ways of adapting membership inference to exploit CP to its fullest.

\subsubsection{Membership inference from first principles.}  
As Carlini et al.~\cite{carlini2022membership} observed, membership inference is an asymmetric  
attack: the attacker is more interested in making a reliable prediction for true positives  
(i.e., predicting “member” when $z \in Z$), rather than true negatives; this is because  
non-members are much more prevalent (they are represented by all the data in the world that is  
not in $Z$).   
A way to exploit this is through  
\emph{shadow models}: the attacker trains multiple models on different ``mock'' datasets,  
labeling their training points as members and held-out points as non-members~\cite{shokri2017membership}.

On the other hand, observe that CP is based, at its core, on a randomness test that  
allows determining if a sample comes from some distribution or not. 
By applying CP to the outputs of shadow models,
the attacker can determine whether a queried point’s behaviour is typical of the training data,
which gives rise to an effective membership inference attack.
Furthermore, CP’s coverage guarantees can lead
to more reliable true-positive detections: if the shadow models data is generated
properly, in such a way that it can be considered to be IID (or exchangeable) with the true
distribution, then CP guarantees $\geq 1-\alpha$ true positives, for any desired significance level
$\alpha$.

\subsubsection{Subset membership inference.}
We propose another variant of membership inference, which is directly inspired from CP:
the attacker gets access to a model trained on $Z$, and to a set of data records $A = \{z_i\}_{i=1}^k$;
their goal is to find the set $S = Z \cap A$ of elements from the challenge set $A$
that are member of the model's training data $Z$;
the attack's success can be measured more finely, by having the attacker output a set
$S' \subseteq A$,
and measuring its precision and recall with respect to the true set $S$.

There are at least two ways of carrying out this attack.
The first is to apply the shadow-based attack, as sketched in the previous paragraph,
to each individual record $z \in A$: a data record is included if, for a certain $\alpha$,
CP judges it to come from the same distribution as other members.
The second way is slightly more involved:
the attacker could set up a CP classification problem, where each data record $z \in A$
is treated as a label, and the CP's prediction set (which can contain any number of
labels from the set $A$) is returned as the attacker's output $S'$.
Further work is likely needed to understand which of the two approaches is more effective.

While this may seem like a minor variation on membership inference attacks, we argue it opens up
an opportunity for real-world attackers, who are often interested in identifying possible multiple
users from large datasets.
For example, an attacker might want to sell information about what data was
used for training a model.
Through this membership inference attack, which we refer to as
\textit{subset membership inference}\footnote{This should not be confused with set membership inference,
where the attacker is given a set of records, all of which are either members or non-members~\cite{hilprecht2019monte}.},
the adversary gets to reduce the space with error guarantees.
We leave these research directions as future work.

\newpage

\section{Traffic analysis attacks}
\label{sec:wf}

Traffic analysis refers to attacks against communication networks aiming to infer
information from the messages in transit.
Typically, one assumes that the network channel is appropriately protected by encryption,
which forces the attacker to look for patterns in the network traffic that leak information
about the communication's content.

For example, consider an instant messaging app protected via end-to-end encryption. An attacker
looking at the encrypted traffic might see the
volume and timing of a user's messages, even though they are unreadable.
If the user consistently   
sends (or receives) large bursts of messages late at night, the attacker can guess
they are
chatting with specific contacts before bed.
Over time, the attacker can build a profile of   
the user's sleep schedule, social activity patterns, and maybe even their closest friends, all   
without reading the actual messages.

The field of traffic analysis is very broad, and new interesting attack vectors are continuously
discovered to this day~\cite{weiss2024your}.
In this section, we explore the use of CP for carrying out so-called
Website Fingerprinting attacks against Tor, an anonymity network,
and then discuss other opportunities that CP offers for traffic analysis attacks in general.

\subsection{\casestudy{} Case Study: Website Fingerprinting}

\begin{figure}
  \centering
  \includegraphics[width=0.7\textwidth]{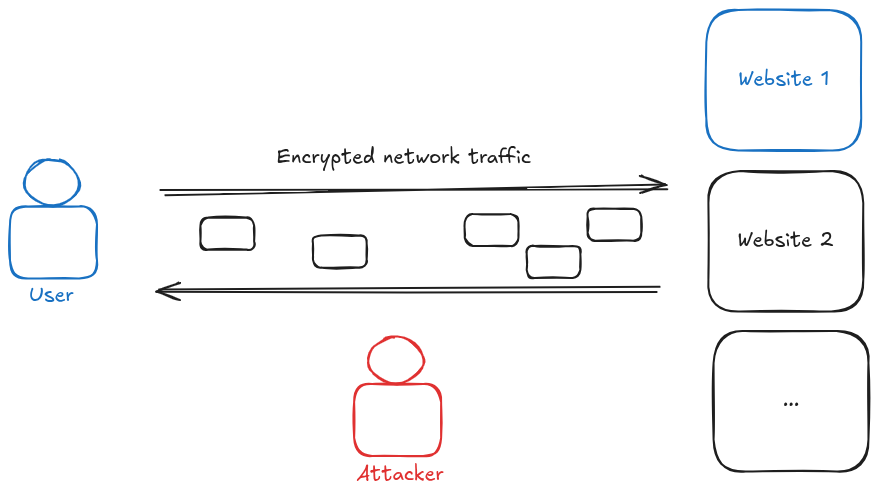}
  \caption{An overview of Website Fingerprinting attacks.
  An attacker monitors the encrypted traffic between a user and the web, and tries to infer which webpages the user is visiting.}
\end{figure}

Anonymity networks, such as Tor~\cite{269582}, enable users to communicate over the Internet in such a way that
an attacker controlling a single point on the network between them cannot figure out who is talking to whom.
Unfortunately, some of these networks are susceptible to traffic analysis attacks, whereby an attacker
who monitors the (encrypted) network traffic between two points infers information about
the communication thanks to patterns in the traffic.

Website Fingerprinting (WF) is possibly one of the most studied traffic analysis attacks
against the Tor network~\cite{cai2012touching,dyer2012peek,johnson2013users,panchenko2016website,hayes2016k,juarez2014critical,cherubin2017bayes,sirinam2018deep,cherubin2022online}.
In WF, an attacker observes network traffic from an encrypted tunnel
(e.g., Tor, a VPN), between a user and the web, and their goal is to infer which webpages
the user is visiting. In traditional WF, the attacker employs an ML classifier to make this inference.
Over the past 15 years, WF attacks have improved substantially,
reaching almost 100\% accuracy in lab conditions.
On the other hand, it is unclear whether WF can be effective in real-world settings~\cite{juarez2014critical},
and empirical measurements suggest WF might only be successful in special conditions,
such as when targeting a restricted number of individuals or webpages~\cite{cherubin2022online}.

\subsubsection{Problem setup and threat model.}
Our setup is deeply tied to a well-known dataset for WF attacks~\cite{sirinam2018deep}.
This dataset contains Tor network traces, each corresponding to a user loading one of
95 websites, 1,000 traces per website.
This simulates the case where an attacker is interested in determining whether a user
visits one of these websites by looking at their Tor network traffic;
we henceforth refer to this set of these websites as
the \textit{monitored} websites.
The dataset also includes an open world (OW) set, containing 40,716 traces from websites
that are not in the monitored set.
This reflects the fact that, in practice, a user rarely visits a predictable set of websites;
therefore, while the attacker is interested in predicting whether they visit a monitored website,
the attacker also needs to discern them from visits to unmonitored (OW) websites~\cite{juarez2014critical}.

We emulate a WF attack via a standard ML classification pipeline:
the classifier is trained on 90\% of the dataset -- the split is done in a stratified manner,
and we evaluate its performance on the rest of the traces.
We employ the attack by Sirinam et al.~\cite{sirinam2018deep}, Deep Fingerprinting (DF),
which uses a convolutional neural network (CNN) to make the prediction.
For the implementation, we adapt the \texttt{wflib} library by Deng et al.~\cite{deng2024wflib}.

\subsubsection{CP for WF.}
We extend CP for carrying out WF attacks as follows.
We use split (or ``inductive'') CP~\cite{papadopoulos2002inductive,papadopoulos2008inductive},
because it can scale to a larger and more complex nonconformity measure
while retaining validity~\cite{vovk2012conditional}.
We use 10\% of the training set as the calibration set, and use the remainder to train
the nonconformity measure (proper training set);
the nonconformity measure is defined as
the complement of the softmax probability scores
(for a desired label) of the CNN trained as per the DF attack on the proper training set.

\subsubsection{Remark: CP with OW traces.}
In traditional WF attacks, one typically needs to train the ML model both on monitored and
unmonitored (i.e., OW) websites; this is to capture the ``unmonitored'' label.
Interestingly, CP lends itself to handling unmonitored (OW) websites by design.
All one needs is to train the nonconformity measure on all monitored traces,
and \textit{ignore the unmonitored websites} during training;
this reflects real-world scenarios, where an attacker only has reliable information on what
are the monitored websites (since they are chosen by the attacker).
Then, when evaluating the attack, CP captures the OW class automatically:
if its prediction set contains no label, this means the trace likely corresponds to
an unmonitored website.
Importantly, CP’s ability to return an empty prediction set when a trace does not match
any monitored website provides a principled way to handle open-world scenarios,
without the need to explicitly model the unmonitored class.

\begin{figure}[htbp]  
    \centering  
    \begin{subfigure}[b]{0.45\linewidth}  
        \centering  
        \includegraphics[width=\linewidth]{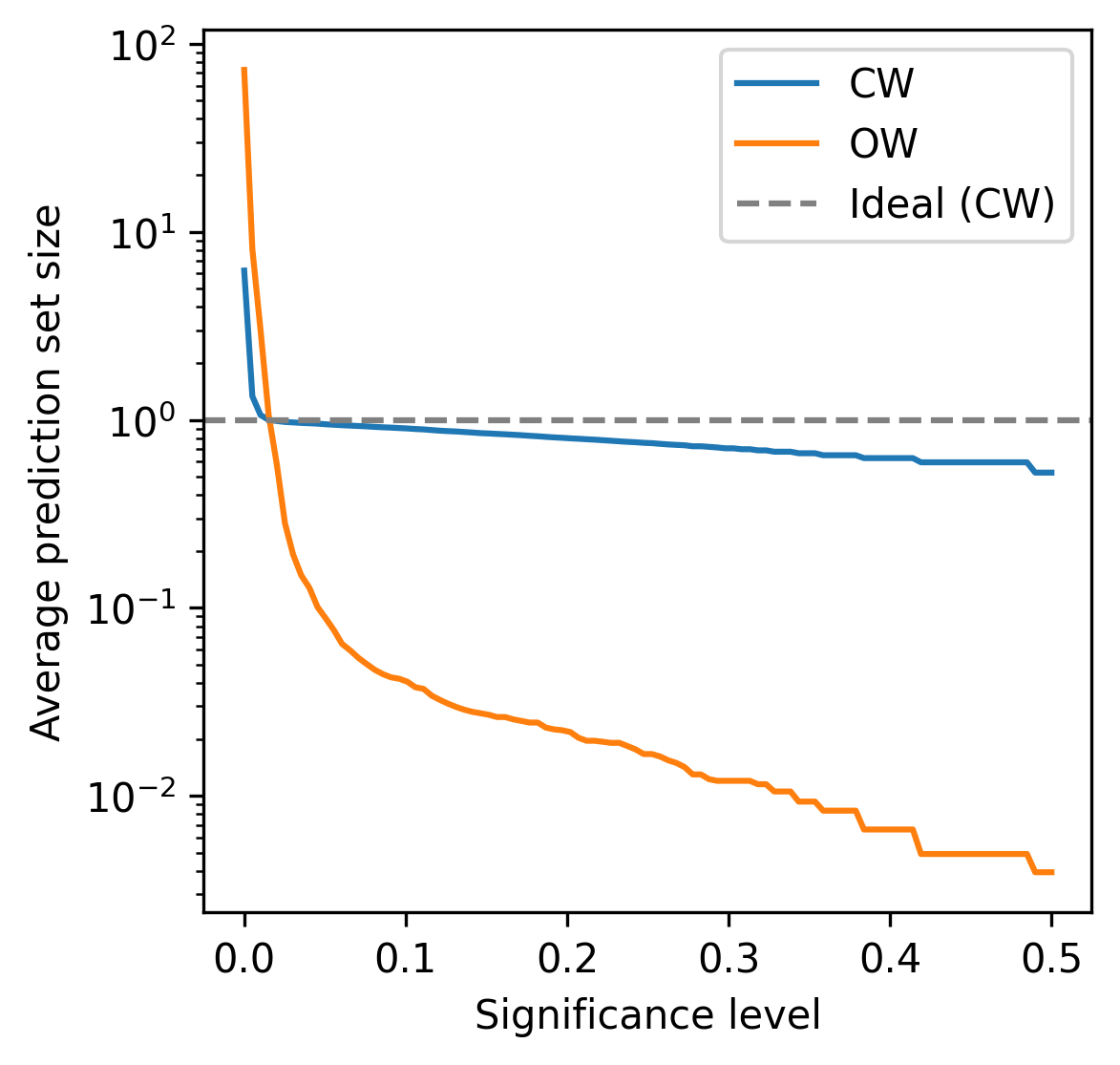}  
        \caption{Prediction set sizes for different significance levels.}  
        \label{fig:cp-predsets}  
    \end{subfigure}  
    \hfill  
    \begin{subfigure}[b]{0.45\linewidth}  
        \centering  
        \includegraphics[width=\linewidth]{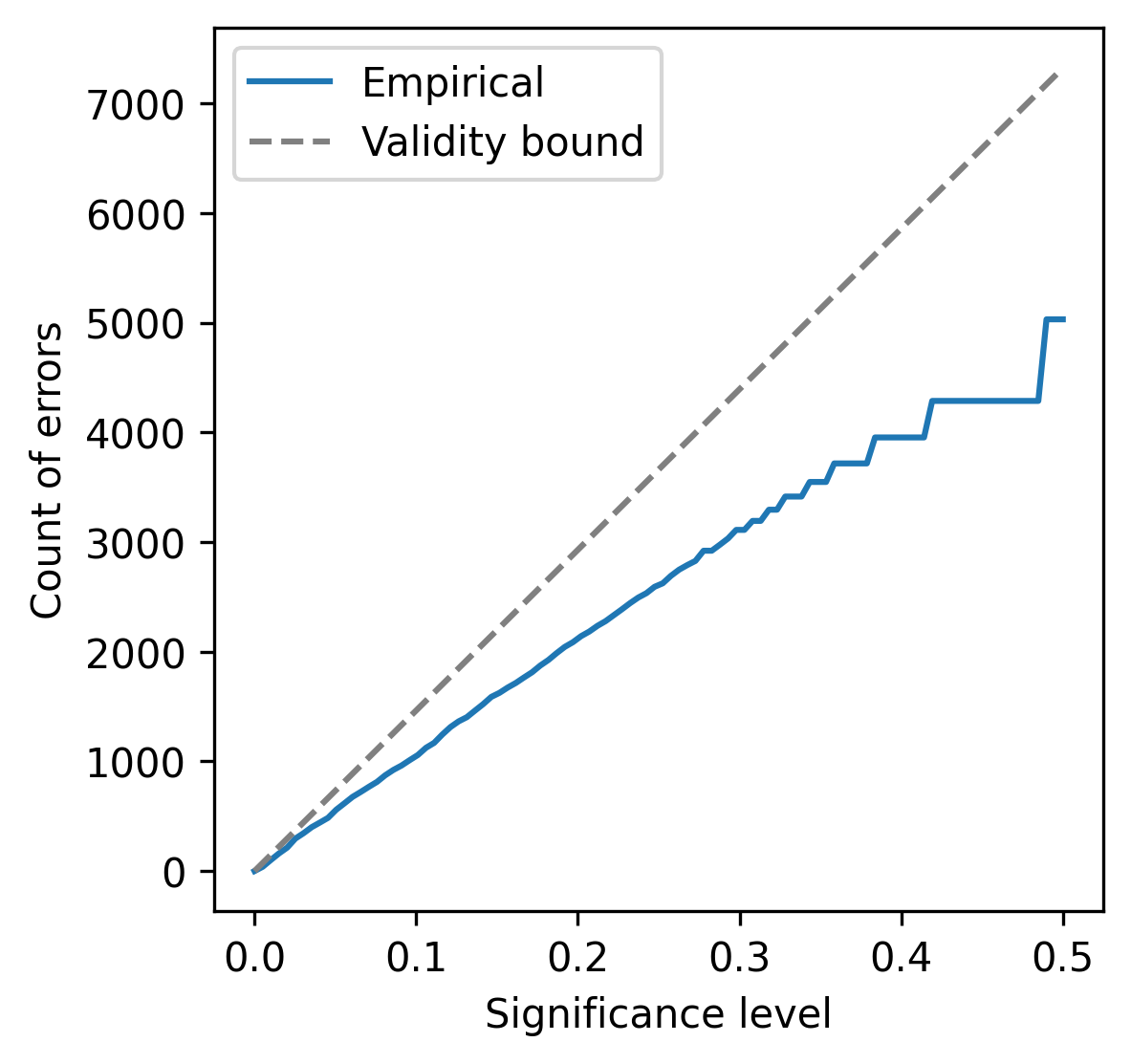}  
        \caption{Miscoverage (error) rates vs. significance levels.}  
        \label{fig:cp-error}  
    \end{subfigure}  
    \caption{WF attack via CP, using DF~\cite{sirinam2018deep} as the basis for its
    nonconformity measure. Results in (a) are shown separately for monitored (CW) and
    unmonitored (OW) websites.}  
    \label{fig:cp-results}  
\end{figure}

\subsubsection{Results.}
\autoref{fig:cp-predsets} reports both the prediction set size and the empirical error
(i.e., number of times the true website is not in the prediction set), as functions of
the significance level $\alpha$.
We measure the prediction set size separately for monitored (CW) and unmonitored (OW)
websites.
Because in this dataset there are no overlaps between websites (e.g., websites
that look very similar), the ideal prediction set size is exactly 1 (i.e., exactly one
label per network trace).
For monitored traces,
we observe that the prediction set size is as expected for reasonably small $\alpha$,
whereas the predictor becomes more conservative as the significance level increases.
This is desirable, and it confirms the effectiveness of CP for carrying out WF attacks.
The prediction set size for unmonitored traces has an interesting behavior:
Whereas negligible values of $\alpha$ force the CP attack to output many websites,
for more common values ($\alpha > 0.015$) the prediction set size becomes smaller than 1;
this is highly informative, as it communicates to the attacker that empty prediction sets
likely correspond to unmonitored traces.

We learn two characteristics, that likely extend beyond WF, and that support CP's suitability
to offensive research:

\begin{tcolorbox}[  
  title=Learnings,  
  colframe=teal!60!white,  
  colback=teal!10,  
  coltitle=black,  
  fonttitle=\bfseries  
]  
\noindent
\textbf{Open world settings.}
In some security attacks we are only interested in a subset of labels,
and for other labels
we may not even have any data to train on.
For example, in WF the attacker is only interested in inferring which websites a user is visiting from a
set of \textit{monitored} websites, which is only a small subset
of all websites. CP offers a great opportunity for attacks with this characteristic:
rather than returning a wrong label,
it will tend to return an empty set;
this informs the attacker that the true label
is not one they are necessarily interested in.

\noindent
\noindent\textbf{Multiple choices are correct.}  
Similarly, there are attacks where multiple labels are acceptable.  
In the WF case, it can be that many websites are indistinguishable in terms of  
network traffic (e.g., their pages' content is very similar), so CP will naturally return many labels in the prediction set.  
For the attacker, this result conveys crucial information:  
if the traffic signature is ambiguous, the set of possible websites remains large,  
and CP captures this uncertainty explicitly.  
This can be valuable in practice, prompting the attacker to collect more data  
or apply additional analyses to further refine which monitored website is relevant. 
\end{tcolorbox}

\subsubsection{Limitations, and how CP can help.}
WF research has received multiple critiques over the years, suggesting that successful
attacks in lab conditions (e.g., network data collected from a single machine, with the
same browser, during the same period of time) may not succeed in the real world~\cite{juarez2014critical,cherubin2022online}.
We posit that tools from the conformal inference literature could help with deploying more practical attacks.
For instance, consider the following problem: websites change over time, and therefore
a WF attack that works today may not work tomorrow because of concept drift.
There are two tools that an attacker can use.
First, they can keep monitoring the size of the prediction set in a CP-based WF attack: this is highly correlated
with the attack's performance, and it can quickly inform the attacker that their
attack needs improving or retraining.
Secondly, they can use methods such as exchangeability martingales to monitor
websites and their network traffic for any distribution changes~\cite{fedorova2012plug,cherubin2018exchangeability,vovk2003testing}.

\subsection{\bluesky{} Research direction: Other attacks to explore}

Traffic analysis is a very broad field, and we foresee CP being useful for many attacks
beyond WF.
We provide two further examples below.

\subsubsection{CP for keystroke analysis.}
A well-known traffic analysis attack against SSH was demonstrated by   
Song et al.~\cite{song2001timing}.   
Since SSH transmits one keystroke per packet, an attacker who measures the time between packets   
can infer what the user is typing, including passwords. Indeed, the time it takes a user to type   
two keys depends on factors such as their distance on the keyboard, and this can be   
exploited to predict the corresponding characters based on timing information.
  
In their original approach, the authors estimated parameters of a hidden Markov model~(HMM)   
from empirical keystroke-timing data. This allowed them to narrow down a list of candidate   
character sequences.
We suggest that, by applying CP to this problem, the attacker   
can explicitly control the error rate (and, indirectly, the size of the candidate sequences)
in a principled manner. For example,
CP-HMM~\cite{cherubin2016hidden} would output a set of likely character sequences,
with a controlled error rate, allowing the attacker to focus on a smaller set of candidates.
In practice, this ensures that the probability of
excluding the correct sequence from the final list is at most~$\alpha$, thereby giving the
adversary a more direct control on the attack's precision.

\subsubsection{Topic prediction from AI chatbot.}
AI chatbots typically stream LLM predicted tokens progressively to the user's browser:
this improves the user experience when interacting with slower LLMs.
On the flip side, this means that each token in an LLM response is sent individually,
which implies that an eavesdropping attacker gets to see the \textit{size} of each token,
even if the content of the actual communication is protected by encryption.
It was recently shown~\cite{weiss2024your} that a sequence of tokens' sizes, corresponding
to an LLM response, might
expose information about the text, such as the topic of the conversation that a user
is having with an LLM.
We foresee CP can be beneficial for this attack. For example, rather than having the attack
predict a single topic, a CP-based attack might make a judgement for each topic independently;
this accounts for cases where multiple topics are discussed in the same conversation with the chatbot.

\section{Discussion}

In this paper, we observed the surprisingly scarce application of CP to offensive security.
In fact, despite a brief suggestion that CP could be used to this end~\cite{cherubin2019black},
we were unable to find any evidence of CP being employed in the field.
Based on our findings, this is remarkable,
as CP is particularly suitable to characteristics
that many security threat models offer, such as:
i) some attacks allow generating virtually unlimited training examples ii) that can be safely
considered to be IID;
furthermore, CP's set predictions (with validity guarantees) are very informative
in cases where iii) the attacker is only concerned with (and can only train on) a subset of all
labels such as open-world scenarios, and iv) where multiple labels are valid predictions
when carrying out an attack.

More in general, we argued that CP matches a quintessential aspect of real world security,
where a set prediction is more effective than a point prediction:
in many cases, attackers are more interested in reducing the set of possible
candidates (e.g., when guessing a password), which they can then explore in subsequent
and more costly attacks (e.g., brute-forcing).
Furthermore, we argued that
CP's validity guarantee on the errors is a useful tool, even for attackers, to calculate the risks
involved when reducing the set of viable candidates.

Overall, we believe that CP has significant potential in the context of offensive security.
While this chapter provided somewhat diverse examples, we suspect that new interesting CP-based
attacks will be devised, and we hope this work can help towards that.

\begin{credits}
\subsubsection{\ackname} 
I thank Andrew Paverd for useful discussion.

\end{credits}
\bibliographystyle{splncs04}
\bibliography{biblio}

\end{document}